\documentclass[twocolumn,showpacs,preprintnumbers,amsmath,amssymb]{revtex4}
\usepackage{graphicx}% Include figure files
\usepackage{dcolumn}% Align table columns on decimal point
\usepackage{bm}% bold math

\begin{document}

\def\ket#1{\langle#1\mid}
\def\bra#1{\mid#1\rangle}

%\preprint{TCMP}

\title{Fermionic concurrence in the extended Hubbard dimer}
\author{Shu-Sa Deng}
\author{Shi-Jian Gu}
%\altaffiliation{Email: sjgu@phy.cuhk.edu.hk\\URL:
%http://www.phystar.net/}
\author{Hai-Qing Lin}
\affiliation{Department of Physics, The Chinese University of Hong
Kong, Hong Kong, China}

\begin{abstract}
In this paper, we introduce and study the fermionic concurrence in
a two-site extended Hubbard model. Its behaviors both at the
ground state and finite temperatures as function of Coulomb
interaction $U$ (on-site) and $V$ (nearest-neighbor) are obtained
analytically and numerically. We also investigate the change of
the concurrence under a nonuniform field, including local
potential and magnetic field, and find that the concurrence can be
modulated by these fields.
\end{abstract}

%03.67.Mn
%Entanglement production, characterization, and manipulation (see
%also 03.65.Ud Entanglement and quantum nonlocality; for
%entanglement in Bose-Einstein condensates, see 03.75.Gg)

%03.65.Ud
%Entanglement and quantum nonlocality (e.g. EPR paradox, Bell's
%inequalities, GHZ states, etc.) (for entanglement production
%in quantum information, see 03.67.Mn; for entanglement in
%Bose-Einstein condensates, see 03.75.Gg)

%71.10.Fd
%Lattice fermion models (Hubbard model, etc.)

\date{\today}
\maketitle

\section{Introduction}

Recently, many efforts have been devoted to the entanglement in
strongly correlated
systems\cite{KMOConnor2001,MCArnesen2001,XWang_PRA_64_012313,VSubrahmanyam04,
SJGurecent,AOsterloh2002,TJOsbornee,Shi,GVidal2003,OFSylju03,SJGu03,JVidal04,
JSchliemann_PRA,PZanardi_PRA_65_042101,JWang03,SJGurecent1},
%GLagmago2002,YSun03,GFZhang2003,FVerstraete04},
in the hope that its non-trivial behavior in these systems may
shed new lights on the understanding of physical phenomena of
condensed matter physics. A typical case is the relation of
entanglement to quantum phase
transition\cite{AOsterloh2002,TJOsbornee,Shi,GVidal2003,OFSylju03,SJGu03,JVidal04}.
For example, Osterloh {\it el al.,}\cite{AOsterloh2002} reported
that the entanglement shows scaling behavior in the vicinity of
quantum phase transition point of the transverse-field Ising
model. Most of previous works are for spin 1/2 systems, where the
degrees of freedom at each site (or qubit) is 2. For these
systems, the entanglement of formation, i.e., the
concurrence\cite{WKWootters98}, is often used as a measure of
pairwise entanglement between two qubits. Such measure is only
valid for $2\otimes 2$ systems. If the degrees of freedom at each
qubit is larger than 2 (for example, the spin 1 system or systems
consisting of fermions with spin), how to quantity the
entanglement of arbitrary entangled state is a challenging issue.
Several studies \cite{APeres96,GVidal02,FVerstraete04,FMintert04}
were devoted to this issue. For example, Mintert {\it et al.}
obtained a lower bound for the concurrence of mixed bipartite
quantum states in arbitrary dimensions. Nevertheless, it is still
very difficult to provide a reliable measure for the pairwise
entanglement of systems with the number of local states larger
than 2. To the best of our knowledge, none of previous work
investigated the pairwise entanglement for systems consisting of
electrons with spin, such as the Hubbard model, although there
were a few works studied the local entanglement of fermionic
models \cite{PZanardi_PRA_65_042101,SJGurecent1,JWang03}.

In this paper, we introduce and study the fermionic concurrence by
using the extended Hubbard dimer as an example. Besides its
importance in exploring many-body correlation in condensed matter
physics, a dimer system also has potential utility in the design
of quantum device\cite{SHill03}. By considering the Pauli's
principle, we first illustrate how to quantify the fermionic
concurrence in the Hubbard model and formulate it in terms of
fermion occupation and correlation so one can easily calculate it.
Then based on the exact solution of the Hubbard dimer, we obtain
the result at the ground state and show that the fermionic
concurrence could be used to distinguish state exhibiting
charge-density correlation from state exhibiting spin-density
correlation. We also study its behavior at finite temperatures
\cite{MCArnesen2001}, and find that it is a decreasing function of
temperature. Moreover, we investigate the behavior of the
concurrence under a nonuniform local potential and magnetic field
\cite{GLagmago2002,YSun03,GFZhang2003}. We find that the
concurrence could be modulated by these local fields. Our work
therefore not only provides a possible way to investigate the
pairwise entanglement in the electronic system, but also enriches
our physical intuition on the systems with both charge and spin
degree of freedom. Some results are instructive for future
experiments.

\section{The model and formulism}
The Hamiltonian of the one-dimensional extended Hubbard model
reads
\begin{eqnarray}\label{eq:Hamiltonian}
H=-t \sum_{\sigma,j,\delta}c^\dagger_{j,\sigma}c_{j+\delta,
\sigma}+U \sum_j n_{j
 \uparrow}n_{j \downarrow}+V\sum_j n_j n_{j+1}.
\end{eqnarray}
where $\sigma=\uparrow,\downarrow;\,j=1,\dots, L; \delta=\pm 1$,
$c^\dagger_{j \sigma}$ and $c_{j \sigma}$ create and annihilate an
electron of spin $\sigma$ at site $j$, respectively, and the
hoping amplitude $t$ is set to unit. At each site, there are four
possible states, $|0\rangle_j,\,|\uparrow\rangle_j,\,
|\downarrow\rangle_j,\,|\uparrow\downarrow\rangle_j$ denoted by $
|\nu\rangle_j,\, \nu=1,2,3,4 $.
 The Hilbert space of
$L$-site system is of $4^L$ dimension, and
$|\nu_1,\,\nu_2\,\cdots\nu_L\rangle= \prod_{j=1}^L|\nu_j
\rangle_j$ are its natural bases. Therefore any state in such a
system can be expressed as a superposition of the above bases. We
consider reduced density matrix $\rho_{jl}={\rm Tr}_{jl}\rho$ of
site $j$ and $l$, where $\rho$ is the thermal density matrix of
the system, and ${\rm Tr}_{jl}$ stands for tracing over all sites
except the freedom of $j$th and $l$th site. Thus $\rho_{jl}$
defines the quantum correlation between site $j$ and site $l$.
However, since there does not exist a well-defined entanglement
measurement for a mixed state of $4\otimes 4$ bipartite systems,
it is impossible to study the entanglement between two sites
exactly. Fortunately, the Hilbert space of site $j$ and $l$ can be
expressed in a direct-product form of electrons with spin up and
spin down, that is, for site $j$ and $l$, we have two subspaces,
one with the bases $|0, 0\rangle, |0, \uparrow\rangle, |\uparrow,
0\rangle, |\uparrow, \uparrow\rangle$, and the other one with $|0,
0\rangle, |0, \downarrow\rangle, |\downarrow, 0\rangle,
|\downarrow, \downarrow\rangle$, respectively. It is therefore
possible to investigate the problem in a given sector with spin up
or spin down separately. For convenience, we restrict our studies
in the subspace of spin up electrons, and the other part can be
simply obtained via the $Z_2$ symmetry. The Hamiltonian
(\ref{eq:Hamiltonian}) possesses U(1)$\times$SU(2) symmetry, i.e.,
it is invariant under the gauge transformation
$c_{j,\sigma}\rightarrow e^{i\theta}c_{j,\sigma}$ and spin
rotation $c_{j,\sigma}\rightarrow U_{\sigma\delta}c_{j,\delta}$,
which manifest the charge and spin conservation. The latter
implies the absence of coherent superposition between
configurations with different eigenvalues of
$n_{j,\uparrow}+{n_{j+1,\uparrow}}$. Thus the reduced density of
electrons with spin-up on two sites has the form
\begin{eqnarray}
\rho_{j,l}= \begin{pmatrix}
  u^+ & 0 & 0 & 0 \\
  0 & w_1 & z & 0 \\
  0 & z^* & w_2 & 0 \\
  0 & 0 & 0 & u^-
\end{pmatrix}
\label{eq:reducemat}
\end{eqnarray}
in the standard bases $|0, 0\rangle, |0, \uparrow\rangle,
|\uparrow, 0\rangle, |\uparrow, \uparrow\rangle$. Elements in the
density matrix $\rho_{jl}$ are related to single occupation and
correlations between the two sites,
\begin{eqnarray}
&&z=\langle c_{j,\uparrow}^\dagger c_{l,\uparrow}\rangle,\nonumber \\
&&u^-=\langle n_{j,\uparrow}n_{l,\uparrow}\rangle, \nonumber \\
&&w_1=w_2=\langle n_{j,\uparrow}\rangle-\langle n_{j,\uparrow}
n_{j+1, \uparrow}\rangle, \nonumber \\
&&u^+=1-2\langle n_{j,\uparrow}\rangle+\langle n_{j,\uparrow}
n_{j+1, \uparrow}\rangle.
\end{eqnarray}
where $\langle\rangle$ denotes the expectation value of the
corresponding operator.

We use the concurrence as a measure of entanglement for such
two-qubit system. It is defined in terms of the spectrum of the
matrix $\rho_{jl}\tilde{\rho}_{jl}$\cite{WKWootters98} where
$\tilde{\rho}_{jl}=\sigma^y_j \otimes\sigma_l^y \rho_{jl}^*
\sigma_j^y \otimes\sigma^y_l$. Precisely, if $\lambda_i$s are
eigenvalues of $\rho_{jl}\tilde{\rho}_{jl}$ and
$\lambda_1\geq\lambda_2\geq\lambda_3\geq\lambda_4$, the
concurrence can then be calculated as
\begin{eqnarray}
C=\max\left[0,
\sqrt{\lambda_1}-\sqrt{\lambda_2}-\sqrt{\lambda_3}-\sqrt{\lambda_4}\right].
\end{eqnarray}
Since there exists a monotonous relation between the concurrence
and the entanglement of formation $E_f$,
$E_f=-x\log_2x-(1-x)\log_2(1-x)$, where $x=1/2+\sqrt{1-C^2}/2$
\cite{WKWootters98}, we will hereafter use the concurrence instead
of entanglement of formation in our study. From Eq.
(\ref{eq:reducemat}), the fermionic concurrence can be calculated
as
\begin{eqnarray}
C= 2\max\left\{|\langle c_{j,\uparrow}^\dagger
c_{l,\uparrow}\rangle|-\langle n_{j,\uparrow} n_{l,
\uparrow}\rangle,\; 0\right\}
\end{eqnarray}

\section{Hubbard dimer with two electrons}

In this section, we consider a model which consists of two sites
and two electrons, because not only can it be exactly solved, but
also it gives us a clear physical picture. The Hamiltonian for the
dimer reads
\begin{eqnarray}
H&=&-2\sum_{\sigma}(c^\dagger_{1,\sigma}c_{2,\sigma}
  +c^\dagger_{2,\sigma}c_{1,\sigma}) \nonumber \\ &&
  +U(n_{1,\uparrow}n_{1,\downarrow}+n_{2,\uparrow}n_{2,\downarrow})
  +2Vn_1 n_2. \label{eq:Hamiltoina_twosite}
\end{eqnarray}
In the standard bases $|\uparrow, \downarrow\rangle, |\downarrow,
\uparrow \rangle, |\uparrow\downarrow, 0\rangle, |0,
\uparrow\downarrow\rangle$ of the reduced subspace with zero
magnetization: $S^z=0$, $H$ is a $4\times 4$ matrix
\begin{eqnarray}
H=\begin{pmatrix}
  U & 0 & -2 & -2 \\
  0 & U & -2 & -2 \\
  -2 & -2 & 2V & 0 \\
  -2 & -2 & 0 & 2V
\end{pmatrix},
\end{eqnarray}
and we can easily obtain its eigenfunction
\begin{eqnarray}
&&\Psi_1=\frac{\sin\theta}{\sqrt{2}}\left[|\uparrow\downarrow, 0
\rangle + |0, \uparrow\downarrow\rangle\right] +
\frac{\cos\theta}{\sqrt{2}}\left[|\uparrow, \downarrow\rangle +
|\downarrow, \uparrow\rangle\right] \nonumber\\
&&\Psi_2=\frac{\cos\theta}{\sqrt{2}}\left[|\uparrow\downarrow, 0
\rangle + |0, \uparrow\downarrow\rangle\right] -
\frac{\sin\theta}{\sqrt{2}}\left[|\uparrow, \downarrow\rangle +
|\downarrow, \uparrow\rangle\right]\nonumber \\
&&\Psi_3=\frac{1}{\sqrt{2}}\left[|\uparrow\downarrow, 0 \rangle -
|0, \uparrow\downarrow\rangle\right] \nonumber \\
&&\Psi_4=\frac{1}{\sqrt{2}}\left[|\uparrow, \downarrow\rangle -
|\downarrow, \uparrow\rangle\right]\label{eq:GSSS}
\end{eqnarray}
where $\theta$ is defined as
$$
\tan\theta=\frac{4}{(U/2-V)+\sqrt{(U/2-V)^2+16}}.
$$
The corresponding eigenvalues are
\begin{eqnarray}
&&E_1=\frac{1}{2}(U+2V-\sqrt{64+(U-2V)^2}), \nonumber \\
&&E_2=\frac{1}{2}(U+2V+\sqrt{64+(U-2V)^2}), \nonumber \\
&&E_3=U, \nonumber \\
&&E_4=2V.
\end{eqnarray}
>From the ground-state wavefunction, we have
\begin{eqnarray}
\langle c^\dagger_{1, \uparrow} c_{2, \uparrow}\rangle &=&
\sin\theta \cos\theta \nonumber \\
\langle n_{1,\uparrow} n_{2, \uparrow}\rangle &=& 0.
\end{eqnarray}
Thus, the fermionic concurrence takes the form of
\begin{eqnarray}
C=|\sin 2\theta|.
\end{eqnarray}

%%%%%%%%%%%%%%%%%%%%%%%%%%%%%%%%%%%%%%%%%%%%%%%%%%%%%%%%%%%%%%%%%%
% Figure 1: the dependence of Ev on U and V
%%%%%%%%%%%%%%%%%%%%%%%%%%%%%%%%%%%%%%%%%%%%%%%%%%%%%%%%%%%%%%%%%%
\begin{figure}
\includegraphics[width=8.5cm]{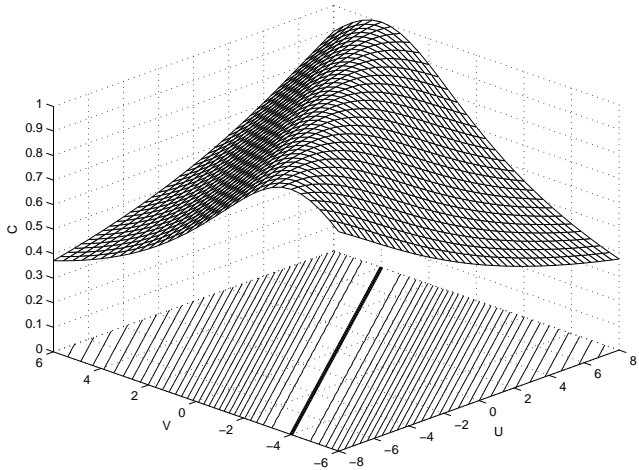}
\caption{The fermionic concurrence of the extended Hubbard dimer
as a function of $U$ and $V$. The curves in the $C=0$ plane
constitutes a contour map, and the thick line denotes the local
extremum. \label{figure_vuc} }
\end{figure}

%%%%%%%%%%%%%%%%%%%%%%%%%%%%%%%%%%%%%%%%%%%%%%%%%%%%%%%%%%%%%%%%%%
% Figure 2: the dependence of Tc on U and V
%%%%%%%%%%%%%%%%%%%%%%%%%%%%%%%%%%%%%%%%%%%%%%%%%%%%%%%%%%%%%%%%%%

\begin{figure}
\includegraphics[width=8.5cm]{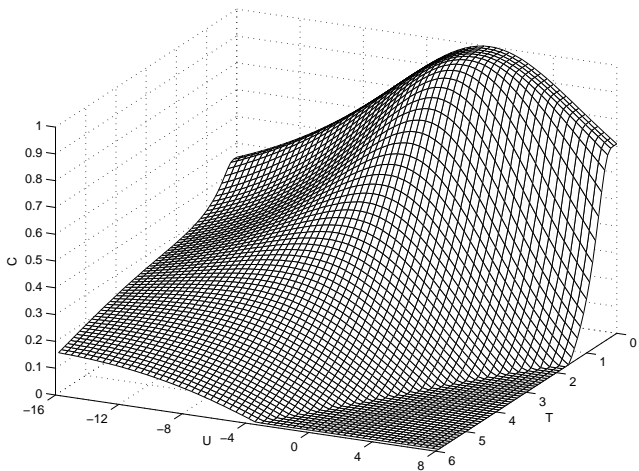}
\caption{The fermionic concurrence of the Hubbard dimer ($V=0$) as
a function of $U$ and $T$.
\label{figure_ve0} }
\end{figure}

\begin{figure}
\includegraphics[width=8.5cm]{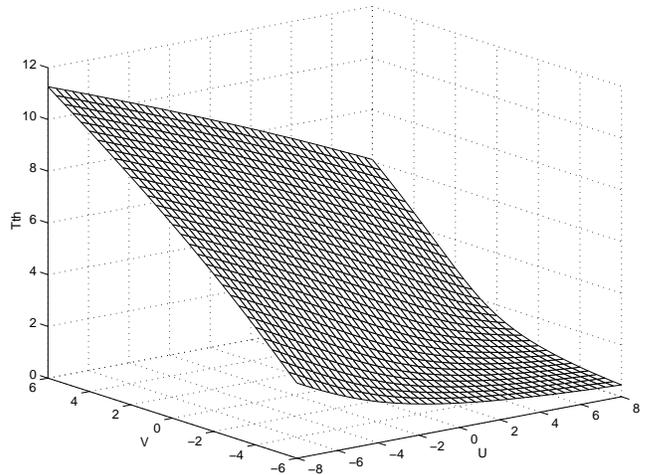}
\caption{The threshold temperature $T_{th}$ of the fermionic
concurrence as a function of $U$ and $V$. \label{figure_tth} }
\end{figure}

Clearly, the ground-state wavefunction simultaneously comprises
the states of the double occupied state $|\uparrow\downarrow, 0
\rangle + |0, \uparrow\downarrow\rangle$ and the single occupied
state $|\uparrow, \downarrow\rangle + |\downarrow,
\uparrow\rangle$. The magnitude of $\theta$ determines which of
them dominates in the ground state. From its definition, it is
easy to learn that the line $U=2V$, i.e. $\theta=45^\circ$,
separates the two regions. One region dominated by
$|\uparrow\downarrow, 0 \rangle + |0, \uparrow\downarrow\rangle$
corresponds to the charge-density-wave state for a large system,
and the other one dominated by $|\uparrow, \downarrow\rangle +
|\downarrow, \uparrow\rangle$ corresponds to the spin-density-wave
state. Clearly, the fermionic concurrence reaches a maximum at the
line $V=U/2$, and shows the mirror symmetry along the line $V=U/2$
(See figure \ref{figure_vuc}).

At finite temperatures, the expectation value of a given operator
$O$ is $ \langle O\rangle=\sum_{n}\langle n|O|n\rangle e^{-\beta
E_n}/Z $, where $\beta=1/T$, $Z$ is the partition function, and
$E_n$ denotes the energy spectra. For the present model, we should
also take into account states in the subspace with $S^z=\pm 1$
besides that of $S^z=0$. Then we have two additional states
$\Psi_{5(6)}=|\uparrow,\uparrow\rangle(|\downarrow,
\downarrow\rangle)$ with the same eigenvalue $E_{5(6)}=2V$. Thus
the fermionic concurrence of the Hubbard dimer takes the form,
\begin{eqnarray}
C(T)=\frac{1}{Z}\max\left[|\sin 2\theta(e^{-\beta E_1}-e^{-\beta
E_2 })|-2 e^{-\beta E_5},\, 0\right]. \label{eq:contemp}
\end{eqnarray}
It is not difficult to prove that the above expression is a
decreasing function of $T$, as shown in figure \ref{figure_ve0}
for a specific value of $V=0$.

In the high temperature limit, $T\rightarrow \infty$, the
Boltzmann weight of each eigenstate becomes almost equal, so the
first term in the square brackets of Eq. (\ref{eq:contemp})
becomes negative and we have a vanishing fermionic concurrence.
Hence we expect that there exists a threshold temperature $T_{\rm
th}$ at which the concurrence becomes zero. Precisely, it is
determined by
\begin{eqnarray}
\left|\sin 2\theta(e^{-E_1/T_{\rm th}}-e^{-E_2/T_{\rm th}})\right|
-2 e^{-E_5/T_{\rm th}}=0. \label{eq:tccond}
\end{eqnarray}
Clearly, the hoping-correlation $\langle c_{j,\uparrow}^\dagger
c_{l,\uparrow}\rangle$ arises from the state of $\Psi_1$ and
$\Psi_2$, and charge correlation $\langle n_{1,\uparrow}n_{2,
\uparrow}\rangle$ from $\Psi_5$. Since the Boltzmann weight
$e^{-E_1/T_{\rm th}}$ of $\Psi_1$ is always larger than
$e^{-E_2/T_{\rm th}}$, the term of absolute value is then positive
for finite $T$. However, the term $U+2V$ in the expression of
Boltzmann wight breaks the mirror symmetry of the fermionic
concurrence at the ground-state along the line $U=2V$. For
example, the concurrence in the fourth quadrant of the $U-V$ plane
is more susceptive to thermal fluctuation and might be easily
suppressed to zero than other quadrants. Therefore, in different
quadrants, the effect of thermal fluctuation is quite different.
Then from Eq. (\ref{eq:tccond}), we obtain a slope-like threshold
temperature for the concurrence (see figure \ref{figure_tth}).

\begin{figure}
\includegraphics[width=8.5cm]{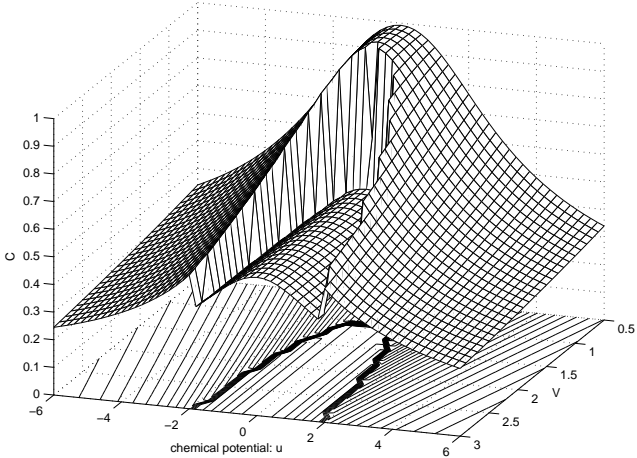}
\caption{The fermionic concurrence as a function of the local
potential $\mu$ ($u$ instead of $\mu$ as an indictor for axes) and
V at the ground-state. Here, $U=2$. \label{figure_vu00} }
\end{figure}

\section{Under a nonuniform field}
In this section, we would like to consider the behavior of the
concurrence under an external field. In general, a uniform
external field, like the local potential or the magnetic field,
suppresses the concurrence since it frustrates the hoping process.
Thus we are only interested in the case with a nonuniform field.
Moreover, differing from the previous investigation where
calculations were done for canonical ensembles, we perform
calculation for grand canonical ensembles. The first case we
consider is to add site energy term to the Hamiltonian
(\ref{eq:Hamiltoina_twosite})
\begin{eqnarray}
H_{\mu}=\mu_1(n_{1,\uparrow}+n_{1,\downarrow})
+\mu_2(n_{2,\uparrow}+n_{2,\downarrow}).
\end{eqnarray}
For simplicity, we set $\mu=\mu_1=-\mu_2$, since the main physics
is not changed under this condition.

\begin{figure}
\includegraphics[width=8.5cm]{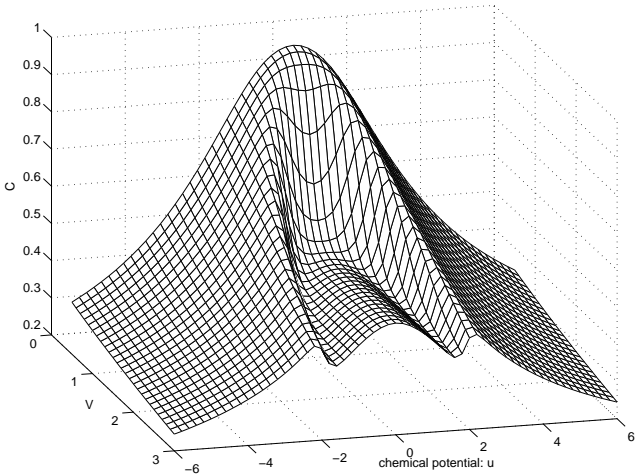}
\caption{The fermionic concurrence as a function of chemical
potential $\mu$ ($u$ instead of $\mu$ as an indictor for axes) and
V at $T=0.1$. Here $U=2$. \label{figure_vu01} }
\end{figure}

\begin{figure}
\includegraphics[width=8.5cm]{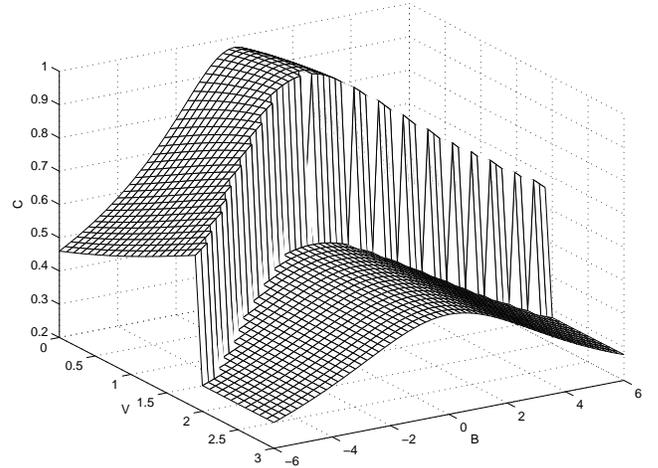}
\caption{The fermionic concurrence as a function of magnetic field
$B$ and V at the ground-state. Here $U=2$. \label{figure_vb00} }
\end{figure}

\begin{figure}
\includegraphics[width=8.5cm]{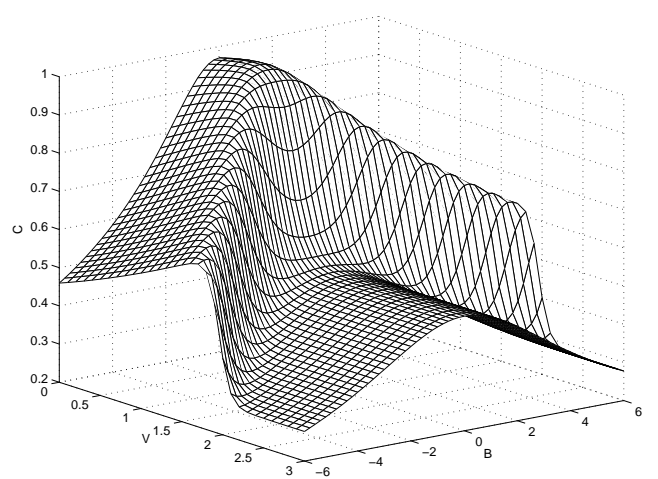}
\caption{The fermionic concurrence as a function of magnetic field
$B$ and V at $T=0.1$. Here $U=2$. \label{figure_vb01} }
\end{figure}

We show the fermionic concurrence as a function of $\mu$ and $V$
as well as its corresponding contour map in the ground-state for a
specific value of $U=2$ in Fig. \ref{figure_vu00}. Clearly the
ground-state phase diagram can be obtained from the contour map.
The area outside the arc is the spin singlet state described in
Eq. (\ref{eq:GSSS}), while the inside area is one-particle state.
Thus the arc in the contour-map is just the level crossing point
for different states. The level-crossing also leads to a jump on
the surface of the concurrence in the $\mu-V$ plane. Since the
experiments are always done at finite temperatures, it is useful
to investigate the effect of thermal fluctuation. We show the
dependence of the fermionic concurrence on $\mu$ and $V$ at
$T=0.1$ in Fig. \ref{figure_vu01} where we observe that the jump
at the level-crossing point in Fig. \ref{figure_vu00} is
eliminated by the thermal fluctuation. We find, at some points,
eg., $U=2$ and $V=1$, the concurrence is small when $\mu=0$. If we
change the value of $\mu$ the thermal concurrence can be increased
substantially. Such a role played by the local potential is
important, because the change of $\mu$ in principle can be
realized by the electric field so it provides a possible way to
control the magnitude of the concurrence.

The second case we consider is to add a Zeeman-like term to the
Hamiltonian (\ref{eq:Hamiltoina_twosite})
\begin{eqnarray}
H_{Zee}=-B_1(n_{1,\uparrow}-n_{1,\downarrow})/2
-B_2(n_{2,\uparrow}-n_{2,\downarrow})/2 .
\end{eqnarray}
For simplicity, we again set $B=B_1=-B_2$. We show the concurrence
as a function of $B$ and $V$ for a specific value of $U=2$ in the
ground state in Fig. \ref{figure_vb00}, and at $T=0.1$ in Fig.
\ref{figure_vb01}, respectively. Similar to the first case, the
magnetic field may also induce the level-crossing in the
ground-state, which may lead to a quantum phase transition. The
induced level-crossing results in a jump on the surface of the
concurrence in the $B-V$ plane. Such a jump is blurred by the
thermal fluctuation, as shown in Fig. \ref{figure_vb01}. Here we
have an interesting phenomena: although the fermionic concurrence
is defined in a single sector, such as for spin-up electrons only,
it can still be modulated by the other sector (spin-down electron)
through magnetic field. For the present case, the magnetic field
does not always decrease the concurrence, it may actually increase
the concurrence in some regions. Therefore, a nonuniform magnetic
field can be in favor of the entanglement of formation between two
qubits.

\section{summary and acknowledgement}

In summary, we first presented a possible way to study the
concurrence as a measure of pairwise entanglement for fermionic
systems. Then we studied its behavior in the Hubbard dimer both at
zero and finite temperatures. It was interesting to observe that
the classification of two different regions, one dominated by the
charge-density correlation $[|\uparrow\downarrow, 0 \rangle + |0,
\uparrow\downarrow\rangle]$ while the other one dominated by the
spin-density correlation $[|\uparrow, \downarrow\rangle +
|\downarrow, \uparrow\rangle]$, is closely related to the behavior
of the fermionic concurrence at the ground state. At finite
temperatures, we also obtained the explicit expression of $C$ for
general $U$ and $V$, and found that it is a decreasing function of
the temperature, which leads to a condition of the threshold
temperature where the concurrence vanishes. We then obtained the
threshold temperature as a function of $U$ and $V$ numerically.
Our method, as we pointed out in the introduction, could be easily
extended to other electronic systems. Results obtained in this
paper help us gain more insight into physical properties of
strongly correlated systems. Moreover, we studied the role of
nonuniform local potential as well as magnetic field for the
extended Hubbard dimer. We found that both of them can be used to
modulate the concurrence. Though our analysis is based on a
specific value of $U=2$, the main physics is the same for other
positive $U$. Finally, we want to point out that although results
obtained in this paper are for a 2-site (dimer) system, their
qualitative features should be the same for big size system.

This work is supported by the Earmarked Grant for Research from
the Research Grants Council (RGC) of the HKSAR, China (Project No
401703). We thank Prof. G. S. Tian for many helpful discussions.

\end{document}